# Tunable Photodetectors Based on 2D Hybrid Structures from Transition Metal Dichalcogenides and Photochromic Molecules

Sewon Park[1], Jaehoon Ji[2], Joakim Andréasson[3], Jeong Ho You[4], and Jong Hyun Choi[1*]

[1] School of Mechanical Engineering, Purdue University, West Lafayette, Indiana 47907, United States
[2] Department of Electrical and Computer Engineering, Princeton University, Princeton, New Jersey, 08544, United States
[3] Department of Chemistry and Chemical Engineering, Chalmers University of Technology, SE-412 96 Gothenburg, Sweden
[4] Department of Mechanical Engineering, University of St. Thomas, St. Paul, Minnesota 55105, United States

[*] Corresponding author: jchoi@purdue.edu

**ABSTRACT**
This article reviews recent progress in two-dimensional (2D) hybrid structures that integrate transition metal dichalcogenides (TMDs) with photochromic molecules for photodetector applications. Atomically thin TMD semiconductors offer strong light-matter interaction, tunable bandgaps, and efficient carrier transport, making them suitable for photodetectors. Photochromic molecules, capable of reversible structural changes in response to external light, can modulate their chemical, optical, and electronic properties. The combinations of various light-adaptive compounds and TMDs provide a versatile platform to explore optically programmable behaviors such as wavelength-selective photoresponse, nonvolatile switching, and multilevel memory characteristics. TMD/photochromic hybrids offer new functional capabilities that are difficult to achieve with either component alone. This mini-review summarizes material properties, interfacial integration strategies, working mechanisms, and representative device demonstrations. We conclude by highlighting future research directions toward the practical implementation of photoresponsive hybrid systems in high-performance adaptive optoelectronics.

# I. INTRODUCTION

The growing demand for high-performance optoelectronic systems has intensified research for novel material platforms and innovative device structures[1]. Photodetectors, which convert light into electrical signals, are essential in a wide range of technologies, including optical communication, biomedical imaging, and environmental monitoring[2, 3]. Conventional photodetectors based on silicon and III-V semiconductors are well-established but face constraints such as limited spectral tunability, lack of flexibility, and integration challenges with emerging device platforms[4, 5]. To address these limitations, researchers have turned to low-dimensional materials that offer tunable optical properties, compact form factors, and compatibility with heterogeneous integration[6, 7].

Transition metal dichalcogenides (TMDs) have emerged as a leading class of two-dimensional (2D) semiconductors due to their unique structural and optoelectronic properties. Atomically thin and mechanically flexible TMDs such as $MoS_2$, $MoSe_2$ and $WSe_2$ exhibit tunable bandgaps and strong excitonic effects, particularly in their monolayer forms[8-11]. These features make them highly suitable for applications in next-generation photodetectors with low power consumption and reduced form factors[6]. Despite this promise, TMD-based photodetectors often suffer from limitations such as fixed spectral responses, slow switching speeds, and a trade-off between responsivity and response time[12-14]. These constraints have motivated researchers to explore hybrid approaches that combine TMDs with other functional materials to unlock new capabilities[15].

One strategy involves integrating TMDs with photochromic molecules, organic compounds that undergo reversible structural changes upon light irradiation, resulting in distinct states with different optical absorption and electronic properties[16-18]. Photochromic molecules alone lack the charge transport and carrier generation capabilities required for efficient photodetection. When integrated with TMDs, they can modulate local electrostatics, influence charge transfer processes, and enable wavelength-selective or reconfigurable photodetector behaviors[19, 20]. To realize such hybrids, strategies like covalent bonding, physisorption, surface anchoring, and van der Waals assembly have been employed, enabling adaptive photoresponse and tunable performance based on molecular and device design[21-24]. Furthermore, parameters such as molecular identity, packing density, and layer thickness can be tuned to precisely control the extent of switching. These materials are poised to contribute to a new class of multifunctional optoelectronic devices in the near future. This mini-review explores the fundamental properties of both TMDs and photochromic molecules relevant to photodetection, followed by a discussion of integration strategies and device structures. Performance metrics, operational mechanisms, and design challenges are also examined, which lays the foundations for future research directions.

# II. Materials Overview: TMDs and Photochromic Molecules

## A. *TMDs: Structure and Properties*

TMDs are layered compounds described by the general formula $MX_2$, where M is a transition metal such as Mo or W, and X is a chalcogen (S, Se, or Te)[7, 25, 26], as shown in Figure 1(a). Each TMD layer comprises a metal atom plane sandwiched between two chalcogen atom planes, forming either trigonal prismatic (2H phase) or octahedral (1T phase) coordination geometries[27]. The semiconducting 2H phase is commonly used in optoelectronic applications, while the metallic 1T and semimetallic 1T' phases, such as in $MoTe_2$, allow phase reconfiguration[14, 27]. Specifically, transitions between these polymorphs can be induced under controlled external stimuli such as thermal annealing, strain, electrostatic gating, or laser irradiation, providing opportunities to tailor electronic properties in device applications[27-29]. These layers are connected by strong in-plane covalent interactions, while adjacent layers are held together by weak van der Waals forces[30, 31]. This structural anisotropy allows for mechanical or chemical exfoliation of bulk crystals into monolayer or few-layer nanosheets[32].

The structural features of TMDs strongly influence their electronic and optical behaviors. Many TMDs are semiconductors with a bandgap ranging from 1 to 2 eV, which can shift from indirect to direct in the monolayer limit[33]. These characteristics allow efficient light-matter interactions, making them well-suited for applications in photodetectors, light-emitting diodes (LEDs), and optical switches [34-36]. The 2D nature also leads to strong exciton binding energies due to weak dielectric screening. Importantly, their electronic properties can be further modulated by intrinsic or extrinsic defects. Sulfur vacancies represent the most common point defects in $MoS_2$ monolayers (Figure 1(b)), serving as chemically active sites that facilitate molecular functionalization or catalytic activity[37, 38]. In the context of photodetector design, such vacancies can act as trap states that prolong carrier lifetimes, thereby enhancing photoconductive gain but potentially reducing response speed. As a result, defect engineering techniques such as controlled doping, plasma treatment, or chemical passivation have been explored to balance sensitivity and temporal resolution in TMD-based optoelectronic devices[39, 40].

Reliable and scalable synthesis of 2D TMDs is essential for their practical deployment in devices. Early studies primarily used mechanical exfoliation or solution-phase sonication to obtain monolayers for fundamental studies due to weak interlayer van der Waals forces and strong in-plane covalent chemistry (Figure 1(c))[7, 21, 32, 41]. In contrast, chemical vapor deposition (CVD) has emerged as a promising synthesis route for producing large-area TMD films with controlled thickness and morphology[42]. This approach enables uniform monolayer growth on substrates such as oxidized silicon, which is critical for device integration. CVD-grown TMDs have been used to fabricate transistors and optoelectronic components with enhanced performance consistency. However, challenges remain, including the formation of grain boundaries, control over crystal orientation, and unwanted defects[43]. Ongoing efforts focus on improving precursor chemistry, substrate selection, and process optimization to enhance material quality and reproducibility. These developments are essential for hybrid device platforms that rely on precise interfaces between TMDs and other functional materials like photochromic compounds.

### B. Photochromic Molecules: Mechanisms and Properties

Photochromic molecules are organic compounds that undergo reversible isomerization between two or more molecular forms in response to specific wavelengths of light, typically UV or visible light[44]. This light-induced change of chemical structure leads to significant modulations in molecular properties such as optical absorption, refractive index, dipole moment, and electronic configuration, forming the basis for their functionality in optoelectronic applications[45-48]. Among the most widely studied classes are azobenzenes, spiropyrans, and diarylethenes illustrated in Figure 2. In azobenzene systems, photochromic changes involve rotation around the N=N double bond, transitioning from a planar trans-form to a bent cis-form. This conformational change alters the molecule's polarity and packing, impacting on its conductivity and interaction with nearby materials. Spiropyrans undergo a ring-opening reaction at the spiro carbon–oxygen bond under UV irradiation, yielding a polar, conjugated merocyanine form. This merocyanine isomer can be reverted to its closed-ring spiropyran structure by exposure to visible light or in a thermal reaction[49-51]. Diarylethenes exhibit an optically reversible 6π-electrocyclization between open and closed-ring isomers, featuring exceptional fatigue resistance and bistability[52, 53]. These molecules are particularly appealing for device integration due to their small size, high synthetic tunability, and robust switching stability over repeated cycles[45].

The diverse functionality of this class of compounds enables their incorporation into various solid-state optoelectronic devices, where they act as molecular switches that respond dynamically to light stimuli. Their reversible switching behavior under ambient conditions also enhances miniaturization and flexible design strategies in modern photonic applications, as well as material properties such as fluorescence, conductivity, and refractive index. In hybrid systems, photoisomerization enables dynamic modulation of the local electrostatic environment, charge transport behavior, and interfacial band alignment. These effects, in turn, influence key photodetector parameters such as threshold voltage, responsivity, and photocurrent dynamics, allowing for reconfigurable and programmable optoelectronic performance in TMD/photochromic devices.

## III. Hybrid TMD/photochromic Photodetectors

### A. TMD Photodetectors: Modes, Performance, and Limitations

Photodetectors based on pristine TMDs leverage the unique optoelectronic properties of 2D materials to enable several operational mechanisms, including photoconductive, photovoltaic, phototransistor, and photoelectrochemical detection[54-56]. In the photoconductive mode, light-induced generation of excess carriers enhances channel conductivity, often with significant gains[57]. Photovoltaic devices, including junctions and Schottky interfaces, separate photogenerated carriers via built-in electric fields and can operate without external bias[58]. Phototransistors modulate channel conductance through gate voltage and incident light, while photoelectrochemical devices exploit photogenerated carriers in redox reactions at active interfaces[36, 59]. More recently, photothermoelectric devices have been developed, converting absorbed light into heat, which generates a photoresponse through temperature gradients and the Seebeck effect[60]. Building on these various operational mechanisms, photodetector architectures have evolved from simple two-terminal layouts to more advanced configurations, such as vertical heterostructures, lateral p–n junctions, and gate-tunable devices[61-63]. While such strategies are also employed in conventional semiconductors, 2D TMDs offer specific structural and electronic features, including their atomically thin geometry, electrostatic tunability, and compatibility with van der Waals integration. These attributes facilitate the development of compact, reconfigurable, and multifunctional device platforms, making TMDs a compelling platform for exploring advanced photodetectors with improved design flexibility and miniaturization.

TMD-based photodetectors exhibit outstanding optoelectronic performance due to their unique electronic structure and strong light-matter interaction in the monolayer regime. Most TMDs such as $MoS_2$, $WSe_2$, $WS_2$, and $MoSe_2$ possess a direct bandgap in their monolayer form, enabling efficient generation and separation of electron-hole pairs under optical excitation[57, 64-66]. The $MoS_2$-based photodetector shown in Figure 3(a)-(b) exhibits high photoresponsivity (~2200 A/W) under optimized conditions.[57] These results reflect efficient carrier multiplication mechanisms. This is further supported by the strong and reproducible photocurrent response under illumination, as shown in Figure 3(c). $MoS_2$ devices can also show strong gate control and clear n-type behavior with an on/off current ratio over $10^3$ under scanning photocurrent microscopy[60]. $WSe_2$-based systems have demonstrated ultrahigh photoresponsivity up to ~$2.2\times10^6$ A/W, along with stable operation at temperatures as high as 700 °C in air and 1000 °C in vacuum[67]. In addition to high sensitivity, TMD photodetectors offer a degree of spectral tunability, for example, by varying the number of layers or applying mechanical strain, the bandgap can be engineered to detect a broad spectral range from UV to near-infrared[64, 68-70]. To exploit multiple TMDs, heterostructured $MoS_2/WSe_2$ layers have been fabricated using epitaxial growth techniques as shown in Figure 3(d)-(f) using band alignment structures for enhanced photodetection[55, 71]. Moreover, TMDs integrated into optical communication circuits have shown fast switching capabilities, with intrinsic response times down to a few microseconds, enabling high-speed photodetection for data transmission[72].

Despite their strengths, TMD-based photodetectors face key challenges. Their limited spectral response, governed by the intrinsic bandgap, restricts adaptability across applications requiring multispectral detection[73]. Persistent photoconductivity, often driven by long-lived trap states, introduces memory effects and hampers rapid signal recovery[74]. TMD surfaces are highly sensitive to environmental exposure, with adsorbed oxygen and water degrading device performances through charge scattering and instability[75]. Another challenge is the trade-off between device responsivity and speed. Controlling charge traps to boost sensitivity also slows the response by extending carrier lifetimes, making it harder to follow fast signal changes[76, 77]. Strategies to overcome these limitations include heterojunction design for broadened absorption, encapsulation for environmental stability, and phase-engineered contacts or work function tuning to improve carrier injection[78-80]. While these conventional strategies are effective, they involve permanent structural changes. Photochromic hybrids, by contrast, enable reversible, light-controlled tuning of key parameters such as responsivity and conductivity, and allow functionalities like optical memory or wavelength-selective detection that are difficult to achieve with static designs[19].

## B. Integration and Interface Engineering

Integrating TMDs with photochromic molecules creates hybrid systems that can respond to light in tunable and reversible ways. The interface between the TMD and the molecular system plays a crucial role in the overall performance of these hybrids. For such hybrids to function effectively, the interface must be engineered with care. Two widely used approaches to achieve this are noncovalent and covalent functionalization. Noncovalent methods rely on weak physical interactions such as van der Waals forces, π- π stacking, and electrostatic attraction [19, 81, 82]. These interactions preserve the TMD's crystal structure and electronic properties. In contrast, covalent functionalization forms strong chemical bonds at specific reactive sites, often associated with intrinsic defects in the TMD lattice[83, 84].

Noncovalent functionalization is frequently used to integrate photochromic molecules. Their attachment occurs through weak physical interactions rather than permanent bonds with 2D TMD layers[85-87]. For example, spiropyrans have been shown to physically adsorb onto $MoS_2$ monolayers, while maintaining their reversible photoisomerization behavior and influencing the TMD surface's local dipole environment upon UV exposure[85, 88]. Diarylethenes have similarly been immobilized on $MoS_2$ flakes via van der Waals interaction, enabling photoresponsive behavior that modulates optical absorption without disrupting the crystal lattice[89, 90]. In many of these systems, the use of tailored ligands or spacer groups, such as long alkyl chains and ethylene glycol moieties, plays a crucial role in enhancing noncovalent conjugation to TMD surfaces[19, 91]. These groups promote stronger van der Waals or dipole interactions, increase surface affinity, and improve the molecular packing density of the photochromic layer, thereby supporting more robust and uniform film formation. For example, alkoxy or alkyl chains in azobenzene and spiropyran can improve adhesion to TMD flakes by enhancing van der Waals interactions as shown in Figure 4(a) and 4(g)[91, 92]. Likewise, ethylene glycol linkers can offer both flexibility and compatibility in solution-based deposition[88, 93]. These molecular design strategies are particularly important for maintaining reliable switching behavior under ambient or fluidic conditions, as they help reduce issues such as molecular aggregation or desorption during repeated optical cycling or at elevated temperatures[93, 94]. However, the physical adsorption can be thermally unstable or reversible under extreme conditions, limiting their suitability for long-term or high-temperature applications[88, 95, 96]. Nevertheless, with careful ligand engineering and interface design, noncovalent strategies offer a versatile and reversible route to modulate TMD optoelectronics without compromising structural integrity.

Covalent functionalization provides a more robust interface by chemically conjugating molecules to TMDs, typically at reactive defect sites. Native chalcogen vacancies, such as sulfur vacancies in MoS$_2$ or selenium vacancies in WSe$_2$, are key targets for thiol-containing molecules[83, 97-100]. Covalent anchoring at these sites can stabilize hybrid interfaces and, in some cases, restore lattice integrity. In one strategy, thiol molecules can be bonded to sulfur vacancies in MoS$_2$ after ion irradiation, improving both lattice quality and molecular integration[37, 101]. Similar functionalization of WSe$_2$ flakes can show not only stable binding but also modulation of its ambipolar charge transport and photoluminescence (PL)[102]. Beyond defect sites, other covalent reactions such as amide bond formation and click chemistry may be applied to anchor photoactive units like azobenzenes or porphyrins onto TMDs[103-105]. In printed MoS$_2$ devices, dithiolated conjugated molecules have been used to chemically bridge adjacent flakes, significantly enhancing charge mobility, on/off ratios, and switching speed in field-effect transistors (FETs)[100]. While these strategies are well established in general TMD molecular hybrid chemistry, their application to photochromic molecules remains relatively rare. This is due to the delicate balance between stable conjugation and photoswitching capability, which requires careful designs of both the molecular constructs and the experimental conditions. One successful example is the covalent integration of diarylethene into MoS$_2$ via diazonium grafting, which has demonstrated stable binding without compromising the molecule's reversible photoswitching capability[106].

While both noncovalent and covalent strategies have been utilized to integrate photochromic molecules with TMD flakes, they differ in terms of experimental complexity and device stability. Noncovalent functionalization, which relies on weak interactions such as van der Waals forces, is widely favored for its synthetic simplicity and compatibility with reversible molecular switching. In contrast, covalent approaches create more stable chemical bonds at defect sites but often involve multistep reactions, surface activation, or post-treatment, which could result in potential changes that hinder photoisomerization. As a result, covalent functionalization remains less commonly used in TMD/photochromic systems. Nevertheless, continued progress in interfacial chemistry and defect control will enable more robust, facile covalent architectures in future hybrid devices.

## C. Coupling Mechanisms and Device Physics

2D hybrid systems made of TMD flakes and photochromic molecules enable light-driven modulation of electronic and optical properties. The interactions are primarily governed by two distinct coupling mechanisms: dipolar interaction and charge transfer. Dipolar interaction, where the isomerization-induced changes in the dipole moment creates local electric fields that shift the carrier distribution in the TMDs[107, 108]. Spiropyrans are well-studied in this context, as the UV-induced isomerization from the spiro-isomer to the merocyanine isomer implies a dramatic increase in the dipole moment, resulting in electrostatic modulation of the adjacent TMD surface[109]. When arranged in a uniform molecular layer, these dipoles collectively form a vertical field that shifts the band alignment and Fermi level of the TMD without requiring charge exchange[85]. The strength of this modulation depends on factors like dipole orientation, molecular packing, and surface coverage[110, 111]. Figure 4(a)-(c) presents this effect on spiropyran-functionalized MoS$_2$, where photoisomerization leads to a transition from a disordered to an ordered molecular packing, as visualized by scanning tunneling microscopy (STM), highlighting collective dipole reorientation and interfacial modulation[91]. Dipolar coupling offers the advantage of reversibility while avoiding any direct chemical modification of the TMDs[81]. The electric field generated by the dipole layer can persist in the dark and be reset with visible light, allowing light-controlled, non-volatile switching behavior[91, 112]. Additionally, dipolar fields can influence excitonic processes in optoelectronic devices. For instance, enhanced electron density in $MoS_2$ can lead to suppression of light emission through nonradiative recombination or improved electrical conduction in transistor applications[16, 113]. With its optical controllability, non-invasiveness, and compatibility with 2D electronics, dipolar coupling represents a promising approach for constructing tunable device architectures[19].

Charge transfer, by contrast, involves the physical movement of electrons or holes across the molecule-TMD interface, governed by the alignment of molecular frontier orbitals with the electronic bands of the TMD. The interfacial electronic configurations and interactions can be understood in detail by using density functional theory (DFT). The first-principles calculations are central to understanding how molecular photoisomerization alters energy-level alignment at the interface, which is a key determinant of charge transfer process. Most DFT studies employ the generalized gradient approximation (GGA) using the Perdew-Burke-Ernzerhof (PBE) exchange-correlation functional, which balances computational efficiency and reasonable accuracy in predicting electronic band structures[90, 106, 114]. DFT workflows typically begin with full geometry optimization of the hybrid system, followed by band structure and density of states (DOS) calculations to assess how the molecule's frontier orbitals (the highest occupied molecular orbital, HOMO, and lowest unoccupied molecular orbital, LUMO) align with the conduction band minimum (CBM) and valence band maximum (VBM) of the TMD. By comparing the open and closed isomer states of photochromic molecules, such as diarylethenes, DFT reveals whether photoisomerization enables charge donation or extraction. Charge density plots and atomic-level charge calculations reveal how electrons move between the molecule and TMD, while dipole and potential changes show how light-driven switching affects the energy bands[115, 116]. These theoretical insights are crucial for interpreting experimental observations such as PL quenching, threshold voltage shifts, or conductivity modulation[20]. They also offer predictive guidance for designing TMD/molecule pairings that maximize optoelectronic tunability in hybrid photodetectors.

The insights from DFT computation inform and guide the photoswitching experiments. Structural isomerization alters the energy levels of the molecule, shifting the HOMO or LUMO relative to the electronic bands of the TMD. For instance, if the LUMO of the photoswitched isomer falls below the CBM of the adjacent TMD layer, photoexcited electrons may migrate from the flake to the molecule. Conversely, if the HOMO lies above the TMD's VBM, hole injection becomes possible[59, 117]. This interaction changes the carrier density, work function, and threshold voltage in TMD via carrier redistribution[20, 90]. Figure 4(d) illustrates the photoisomerization of diarylethene on a MoSe$_2$ monolayer, enabling controlled changes in molecular orbital energies[20]. The energy band alignment in Figure 4(e) suggests that the closed-ring diarylethene isomer has a LUMO below the MoSe$_2$ CBM, facilitating electron transfer. PL measurements show a clear drop in emission intensity after exposure to UV light (Figure 4(f)), consistent with increased charge separation due to photoinduced electron transfer at the hybrid interface. Unlike dipolar coupling, which induces electrostatic shifts, charge transfer affects the occupation of states in the semiconductor and can lead to more pronounced and lasting changes in electronic properties. The efficiency of this mechanism depends critically on molecular design, the strength of electronic coupling at the interface, and the dimensionality of the TMDs, where strong coupling requires close van der Waals contact and minimal tunneling barriers[107]. Because charge transfer can introduce trapping, hysteresis, or recombination centers, however, its impact on device performance must be carefully managed[115, 116].

Both mechanisms, dipolar coupling and charge transfer, can coexist in the same system as illustrated in Figure 4(g) where both charge transfer and dipolar effects occur upon photoisomerization of azobenzene on MoS$_2$ flakes. Figures 4(h)-(i) show Raman and PL spectra which indicate doping and carrier modulation between the cis- and trans-isomeric states, facilitating photodetection. The dominant effect is dictated by interfacial energetics, molecular conformation, and environmental factors such as substrate dielectric environment and ambient conditions[88, 92]. The precise control of energy level alignment, molecular orientation, and interface quality is thus essential to harness the full potential of hybrid photodetectors[84, 108]. These two coupling modes offer complementary routes to build multifunctional hybrid materials with light-tunable band structure, doping profiles, and electronic response.

### D. Device Performance and Case Studies

The integration of photochromic molecules with TMDs enables hybrid systems whose interfacial properties can be dynamically controlled by light-driven molecular switching. As a result, the devices allow for switchable conductance, programmable memory, and wavelength-dependent detection, all activated by light alone[16, 19]. Through careful design of molecular structure and TMD interface, these hybrid architectures have emerged as powerful platforms for building adaptive, low-power optoelectronic systems[92]. The following case studies examine the implementation of such hybrids using various TMDs, revealing how material choice influences functional outcomes.

Photochromic molecules with well-defined switching behavior and interfacial compatibility are effective in 2D hybrid systems, as they can modulate dipole moments or frontier orbital energies to tune charge transport across TMD interfaces. Depending on their interaction mode, such as electrostatic gating or direct charge transfer, these molecules enable diverse mechanisms for light-triggered modulation in 2D hybrid electronics. In MoS$_2$/azobenzene hybrids, photoisomerization induces dipole-driven electrostatic gating that modulates channel conductance without charge transfer[94, 118]. This mechanism has also been used to demonstrate diode-like switching characteristics with tunable rectification ratio under alternating light exposure[94]. Spiropyran-functionalized MoS$_2$ operates through a similar mechanism. Upon UV irradiation, the spiropyran isomer converts into the merocyanine form, introducing a strong out-of-plane dipole that modulates the MoS$_2$ channel electrostatically. This yields distinct threshold voltage shifts and robust electrical modulations[88, 91, 93]. In contrast, MoS$_2$/diarylethene hybrids engage in isomer-dependent charge transfer. The closed form aligns with energetically with the MoS$_2$ valence band, enabling hole injection upon UV-exposure and enhancing conductivity, while the open isomer does not[90, 106]. This results in distinct PL quenching behavior upon UV irradiation, with recovery under visible light. Notably, this switching behavior exhibits strong thickness dependence as shown in Figure 5(a) and (c), where monolayer MoS$_2$ shows pronounced conductance and PL modulation. In contrast, bi-/tri-layer devices display little to no response. This observation is supported by DFT calculations in Figure 5(b) [90]. This layer-sensitive effect highlights the role of interfacial coupling and offers a new handle for thickness-selective optoelectronic tuning.

In WSe$_2$-based devices, spiropyran molecules enable light-controlled modulation of charge transport through interfacial dipole switching, resulting in over 80% current suppression[113]. When combined with a ferroelectric overlayer, the device exhibits enhanced gating control through cooperative dipolar and polarization effects, allowing precise modulation of carrier transport under optical and electrical stimuli[119]. A similar strategy has been applied to WSe$_2$/diarylethene hybrids, where photoisomerization of the diarylethene molecules enabled threshold voltage shifts exceeding 2 V and stable multilevel switching behavior[115]. The ferroelectric polarization amplified the molecular dipole effect, allowing reversible and programmable control of both electron and hole currents under ambient conditions. In addition, a bicomponent diarylethene blend has achieved simultaneous modulation of both electron and hole currents, with photoisomerization suppressing conductance by up to 97% for holes and 52% for electrons[116]. Figure 5(g) illustrates the device architecture of WSe$_2$ integrated with bicomponent diarylethene, which enables ambipolar modulation of charge transport. Under alternating UV and visible light exposure, the device exhibits stepwise and reversible modulation of hole transport, as shown in Figure 5(e), with five clearly distinguishable current levels. A similar trend is observed for electron transport (Figure 5(f)), where stable multilevel states are achieved via controlled isomerization of the diarylethene layer. Together, these results demonstrate the capability of the WSe$_2$/diarylethene hybrid system for programmable, ambipolar, and light-tunable photodetection.

To consolidate these findings and provide a direct performance comparison across different hybrid systems, Table 1 presents a summary of representative TMD/photochromic devices, including their molecular composition, integration method, dominant coupling mechanisms, and key functional behaviors such as threshold shifts, PL modulation, current rectification, and multilevel switching, offering a comprehensive overview of the design strategies and functional outcomes observed in literature.

The case studies present the remarkable tunability of TMD/photochromic hybrid systems. By selecting appropriate TMD materials and tailoring the interfacial chemistry, a wide range of light-controllable functions can be achieved. The ability to reconfigure device behavior on demand using light only is a powerful tool for advanced optoelectronic platforms[19].

**IV. CONCLUDING REMARKS**

This mini-review introduces hybrid 2D materials that combine TMDs with photochromic molecules to develop optically tunable and reprogrammable photodetectors. TMDs offer strong excitonic effects, direct bandgaps in the monolayer limit, and high carrier mobility, making them suitable for compact and sensitive photodetectors[6, 64, 68]. Despite these advantages, they face intrinsic limitations such as fixed spectral response and slow recovery caused by charge traps, which restrict dynamic functionality and response speed[120]. To address these issues, photochromic molecules have been integrated with TMDs. While photochromic molecules lack intrinsic photodetection capability, their integration with TMDs enables active modulation of device behavior via two distinct mechanisms[16, 18]. In charge transfer, molecular switching alters orbital alignment, enabling carrier exchange between the molecule and TMDs[90, 121]. In dipolar coupling, electric fields from molecular dipoles modulate carrier distribution without direct charge flow[81, 85]. These mechanisms have enabled functions such as threshold voltage shifts, PL quenching, and multilevel switching, as demonstrated in various studies[91, 94, 106, 116, 119].

The exceptional properties of TMD/photochromic hybrids, such as light-triggered, nonvolatile switching and programmable transport behavior, highlight their potential in engineering applications. One notable advantage is that the photochromic organic layer may help mitigate environmental instability by acting as a protective coating for the underlying TMD, provided it maintains sufficient chemical stability and degradation resistance[122]. Nevertheless, several technical challenges remain, including molecular fatigue, surface degradation of TMDs under ambient conditions, and difficulties in achieving large-area integration[45, 75, 123]. These issues must be resolved to realize effective integration of molecular switching with 2D semiconductors.

Among the notable advantages of TMD/photochromic systems is their tunability beyond simple binary switching. The degree of optoelectronic modulation may be adjusted continuously by varying packing identity of photochromic molecules, illumination intensity, and exposure duration. These parameters influence the strength of dipole-induced electrostatic fields or the efficiency of interfacial charge transfer, enabling a range of intermediate electronic states. This continuous controllability, however, has received limited attention in the context of photodetection using 2D hybrid devices. Most reported studies focused on qualitative (e.g., on/off) switching behavior. There remains a lack of systematic investigation into how such tunability affects photodetector performance metrics including responsivity, spectral selectivity, detectivity, external quantum efficiency, gain, response time, and noise-equivalent power. Addressing these fundamental gaps will clarify the role of molecular tunability in photodetection and support the development of practical device strategies.

To effectively overcome the challenges discussed above, future research should focus on translating the unique advantages of these hybrid structures into reliable device-level performance. These include rational molecular design to improve switching contrast and chemical stability, precise interfacial engineering to enhance electronic coupling, and the development of robust integration methods that align with standard fabrication processes. In addition, more systematic DFT studies are crucial for gaining insight into energy level alignment and interfacial charge dynamics, as well as for guiding the discovery of new TMD/photochromic hybrid combinations with optimized optoelectronic performance. With continued progress, TMD/photochromic hybrids could become foundational components in emerging technologies such as logic-in-memory architectures, light-gated transistors, and reconfigurable photonic circuits[81].

**NOTES**
The authors declare no competing financial interest.

**ACKNOWLEDGEMENTS**
This work was funded by the U.S. National Science Foundation under award no. 2151869 and 2151887.

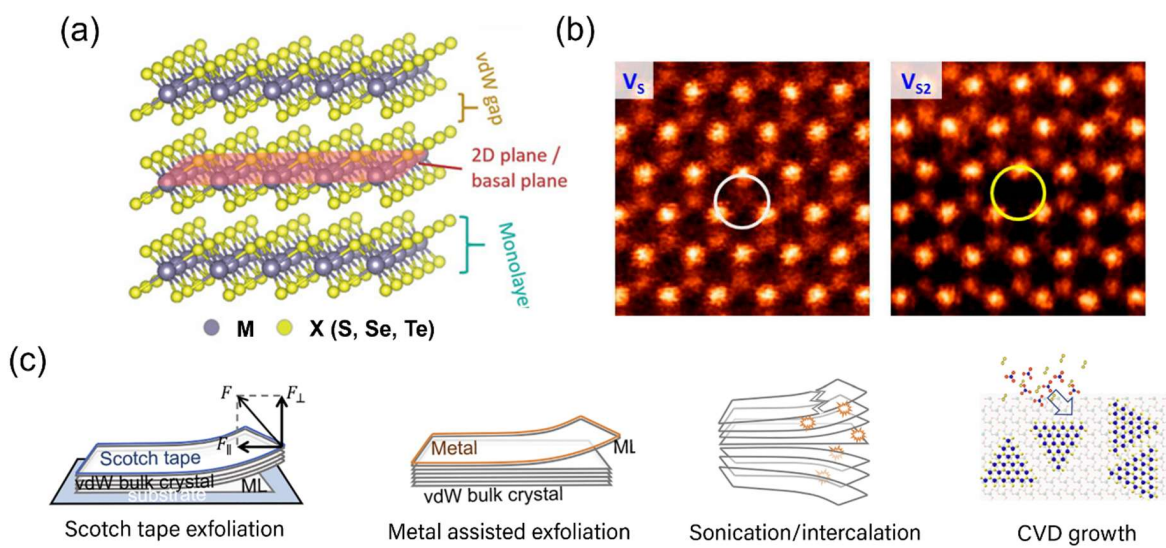

**Figure 1.** (a) Schematic of layered TMD structure showing strong in-plane bonds and weak van der Waals interlayer interactions. (b) Atomic-resolution images of single (left) and double (right) sulfur vacancies in CVD-grown monolayer $MoS_2$. (c) Methods to prepare 2D monolayers: Scotch tape exfoliation, metal-assisted exfoliation, solution-phase sonication/intercalation, and CVD growth. Adapted with permission from refs [26], [43], and [41]. Copyright 2021 Wiley-VCH. Copyright 2013 American Chemical Society. Copyright 2021 Elsevier.

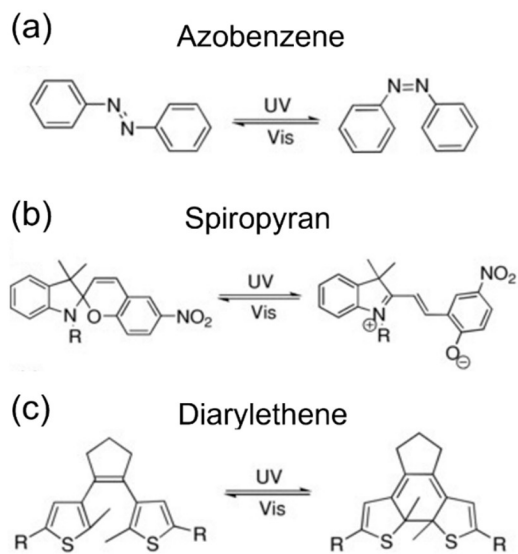

**Figure 2.** Isomerization schemes for a selection of commonly used photochromic classes of compounds. (a) Azobenzene (trans-cis isomerization), (b) spiropyran (ring closing/opening), and (c) diarylethene (ring closing/opening). Reproduced with permission from ref [45]. Copyright 2012 Wiley-VCH.

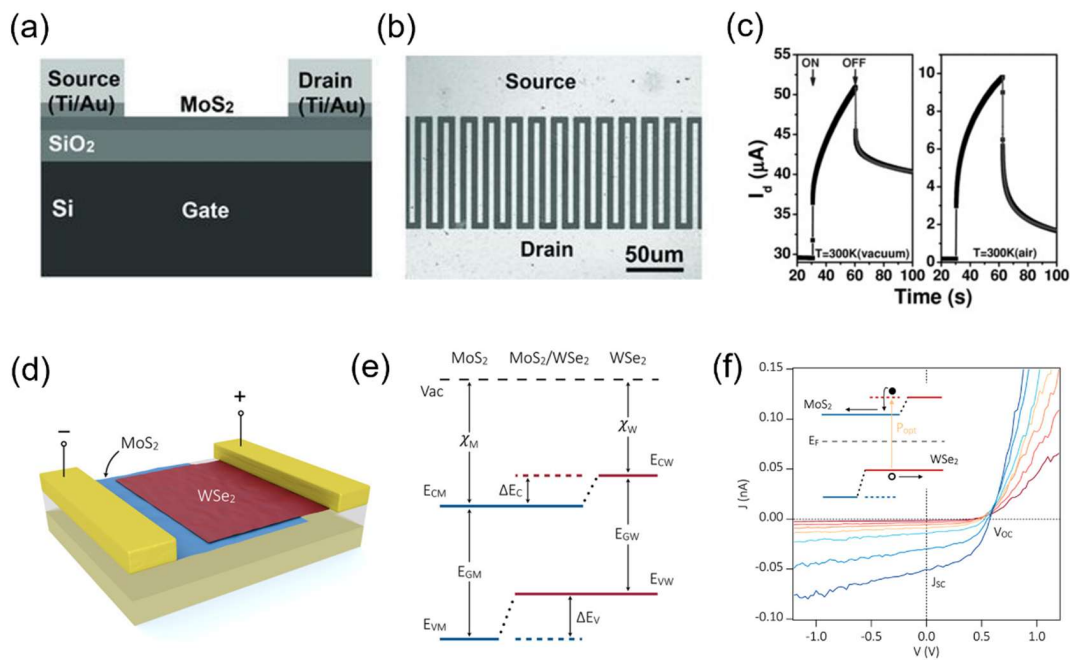

**Figure 3.** (a)-(c) Monolayer MoS$_2$ photodetector. (a) Schematic illustration. (b) Top-view optical microscopy image of the device with source and drain electrodes. (c) Time-resolved photocurrent measurements showing current responses at 300 K in vacuum (left) and air (right). (d)-(f) Photodetector made of MoS$_2$/WSe$_2$ heterostructure. (d) Schematic diagram. (e) Type-II band alignment at the MoS$_2$/WSe$_2$ interface, enabling charge separation with electrons in MoS$_2$ and holes in WSe$_2$. (f) Current-voltage plot driven by the band alignment. Adapted with permission from refs [57] and [55]. Copyright 2013 Wiley-VCH. Copyright 2014 American Chemical Society.

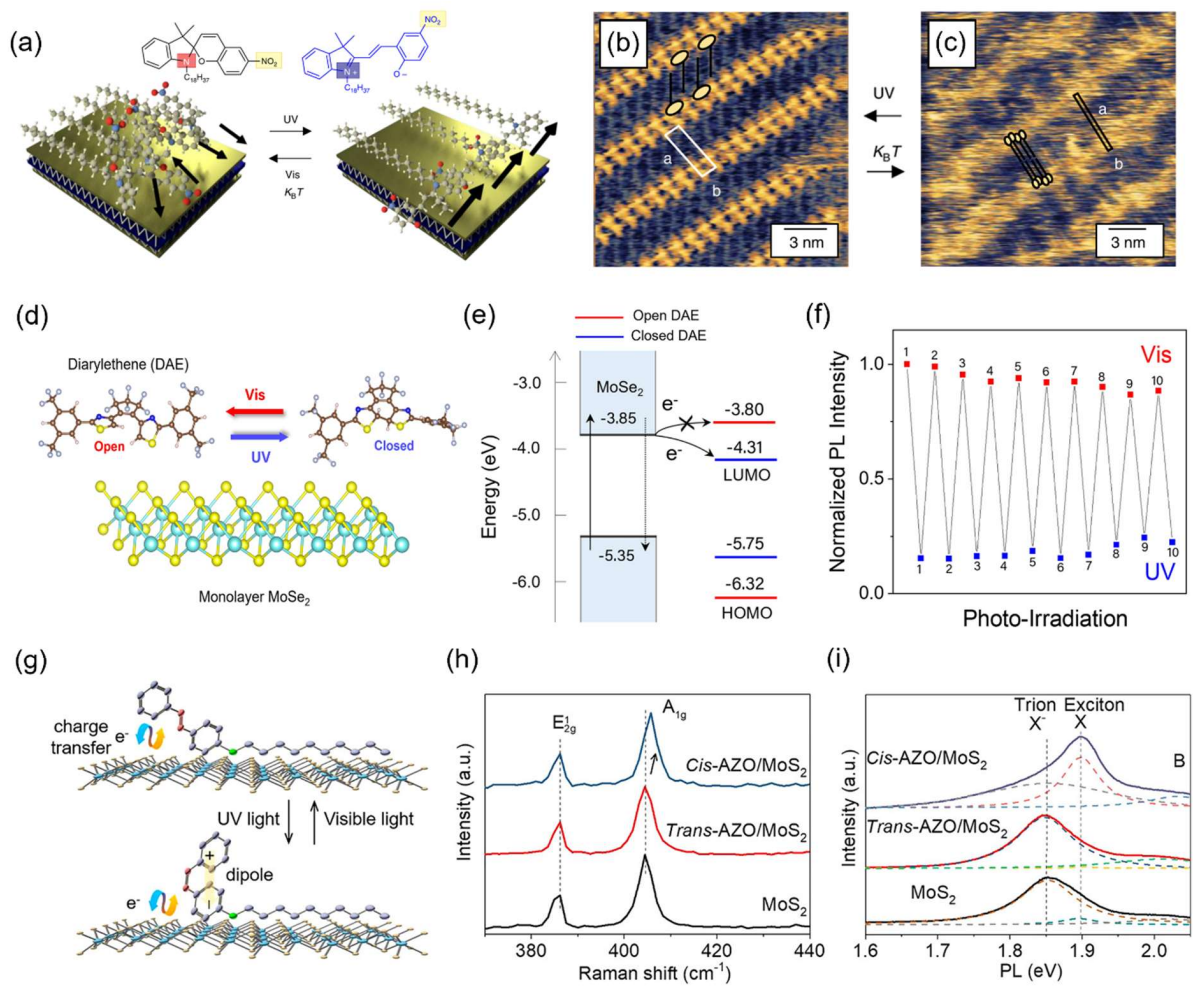

**Figure 4.** (a)-(c) MoS$_2$/spiropyran photodetector. (a) Reversible photo-switching of spiropyran to merocyanine on MoS$_2$, altering dipole orientation and substrate energetics under UV/visible light. (b) Scanning tunneling microscopy (STM) image after UV irradiation shows ordered merocyanine domains with highly ordered lamellae. (c) STM image after thermal relaxation reveals the re-formation of disordered spiropyran lamellae with mobile, randomly oriented head-groups. (d)-(f) MoSe$_2$/diarylethene (DAE) hybrid. (d) Diagram showing the reversible transformation between open and closed forms of diarylethene under visible and UV light on a MoSe$_2$ monolayer. (e) Energy band alignment between MoSe$_2$ and diarylethene isomers. The closed form has a LUMO level lower than the CBM of monolayer MoSe$_2$, allowing for photoinduced electron transfer, while the open form does not. (f) PL intensity plot over alternating UV/visible irradiation cycles, demonstrating reversible and stable switching behavior. (g)-(i) MoS$_2$/azobenzene system. (g) Schematic of dipole-driven charge transfer in MoS$_2$ with azobenzene hybrids, modulated by photoisomerization. (h) Raman spectra showing isomer-dependent doping effects in MoS$_2$. (i) PL spectra showing shifts in trion and exciton peaks (~20 meV) due to interfacial interactions. Adapted with permission from refs [91], [20] and [92]. Copyright 2018 Springer Nature. Copyright 2024 Springer Nature. Copyright 2019 American Chemical Society.

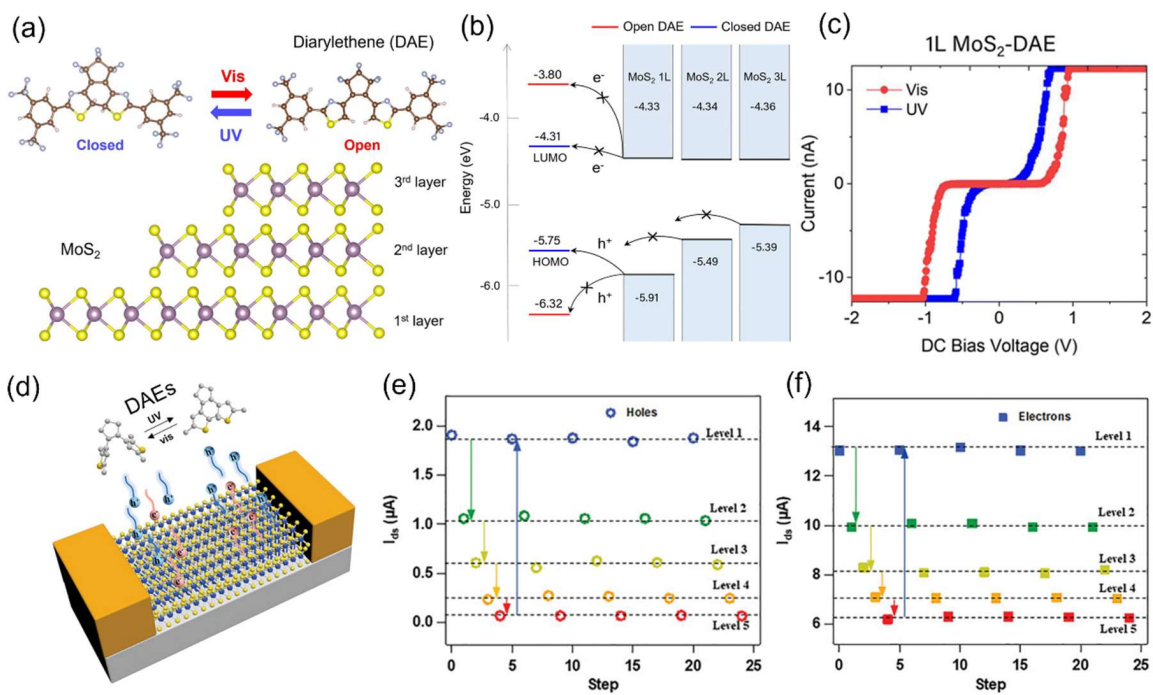

**Figure 5.** (a)-(c) MoS$_2$/diarylethene (DAE) photodetector. (a) Schematic of a layer-dependent MoS$_2$ photodetector functionalized with diarylethene molecules. (b) Energy level diagram showing HOMOs and LUMOs of two DAE isomers aligned with the valence and conduction bands of 1L ~ 3L MoS$_2$, indicating layer-dependent photoswitchable hole transfer. (c) Conductive AFM measurements showing reversible current modulation in monolayer MoS$_2$ upon light-induced isomerization. (d)-(f) WSe$_2$/bicomponent diarylethene hybrids. (d) Illustration of optical modulation of ambipolar charge transport in WSe$_2$ using a diarylethene layer. Optical modulation of drain current, showing five discrete current levels for (e) holes and (f) electrons, is achieved by varying the UV exposure time (0, 10, 20, 40, and 80 sec). Adapted with permission from refs [90] and [116]. Copyright 2025 Royal Society of Chemistry. Copyright 2020 Wiley-VCH.

Table 1. Representative TMD/Photochromic Hybrid Devices

| TMD | Photochromic Molecule | Interfacial Chemistry | Major Coupling Mechanism | Note | Reference |
|---|---|---|---|---|---|
| $MoS_2$ | Azobenzene | Noncovalent | DI | CT dominant in hybrid FETs | 94 |
| $MoS_2$ | Azobenzene | Noncovalent | DI | Doping and work function modulation | 92 |
| $MoS_2$ | Azobenzene | Noncovalent | DI | Photoswitchable diode effect | 118 |
| $MoS_2$ | Azobenzene | Noncovalent | DI | PL modulation and Fermi level shift | 124 |
| $MoS_2$ | Spiropyran | Noncovalent | DI | Threshold voltage shift | 88 |
| $MoS_2$ | Spiropyran | Noncovalent | DI | Collective switching in superlattices | 91 |
| $MoS_2$ | Spiropyran | Noncovalent | DI | Humidity-dependent interaction | 93 |
| $MoS_2$ | Diarylethene | Noncovalent | CT | Electrostatic interaction | 106 |
| $MoS_2$ | Diarylethene | Covalent | CT | Covalent C-S bond via diazonium | 106 |
| $MoS_2$ | Diarylethene | Noncovalent | CT | Thickness-dependent photoresponse | 90 |
| $WSe_2$ | Spiropyran | Noncovalent | DI + CT | Light, electric, heat control | 113 |
| $WSe_2$ | Diarylethene | Noncovalent | CT | Photoinduced Fermi level shift | 115 |
| $WSe_2$ | Diarylethene | Noncovalent | CT | Ambipolar current modulation | 116 |
| $WSe_2$ | Diarylethene | Noncovalent | CT | Light, electric, polarization control | 119 |
| $MoSe_2$ | Diarylethene | Noncovalent | CT | PL modulation | 20 |

Notes: DI, dipole interaction; CT, charge transfer